\documentclass[useAMS,usenatbib]{mn2e}
\usepackage{epsf}

\title[Using distant globular clusters as a test for gravitational theories]{Using distant globular clusters as a test
for gravitational theories}
\author[H. Baumgardt, E. K. Grebel and P. Kroupa]{H. Baumgardt$^{1}$\thanks{e-mail:
holger@astro.uni-bonn.de (HB); grebel@astro.unibas.ch (EKG); pavel@astro.uni-bonn.de (PK)}, 
E. K. Grebel$^2$\footnotemark[1] and P. Kroupa$^{1}$\footnotemark[1]\\
$^{1}$Sternwarte, University of Bonn, Auf dem H\"ugel 71, 53121 Bonn, Germany\\
$^{2}$Astronomical Institute, University of Basel, Venusstrasse 7, CH-4102 Binningen, Switzerland\\
$^{3}$Sternwarte, University of Bonn, Auf dem H\"ugel 71, 53121 Bonn, Germany}
\begin{document}

\date{Accepted ????. Received ?????; in original form ?????}

\pagerange{\pageref{firstpage}--\pageref{lastpage}} \pubyear{2004}

\maketitle

\label{firstpage}

\begin{abstract}
We propose to determine the stellar velocity dispersions of globular clusters
in the outer halo of the Milky Way in order to decide whether 
the dynamics of the universe on large scales is governed by dark matter 
or modified Newtonian dynamics (MOND). We show that for a number of
Galactic globular clusters, both the internal and the external accelerations
are significantly below the critical acceleration parameter $a_0$ of MOND.
This leads to velocity dispersions in case of MOND which exceed their Newtonian counterparts
by up to a factor of 3, providing a stringent test for MOND.
Alternatively, in case high velocity dispersions are found, these would
provide the first evidence that globular clusters are dark matter dominated.
\end{abstract}


\section{Introduction}

Dark matter is now generally believed to be the dominating mass component of the 
universe, starting with the discovery of \citet{z33}
that the speed of galaxies in the Coma cluster is too large to keep them 
gravitationally bound unless they are much heavier than one would 
estimate on the basis of visible matter alone.
Although the currently favoured $\Lambda$CDM model has proven to be remarkably
successful on large scales \citep{sp03}, high-resolution N-body
simulations are still in contradiction with observations on subgalactic scales
where they predict orders of magnitude more substructure than what is seen \citep{m99, k99}
and also a wrong spatial distribution of the subhalos \citep{ktb04}.
Additional arguments regarding supporting and contradictory
observations of structure formation in CDM models are presented in \citet{ggh03} and \citet{gg04}.
The above discrepancies might be resolved if more realistic simulations 
that can better treat the dynamics
of the interstellar gas and feedback processes 
become available, or they could show that our current understanding
of cosmology and large scale structure formation is still missing important ingredients.

An alternative to the dark matter hypothesis could be
Modified Newtonian Dynamics (MOND), which was proposed by \citet{m83a, m83b}
and \citet{bm84} as a way to explain the rotation curves of
galaxies without the need to assume large amounts of otherwise unseen
dark mass in the outer parts of galaxies. According to MOND, Newtonian dynamics
breaks down in the limit of very weak accelerations, and the acceleration $\vec{a}$
experienced by a particle is given by the following (heuristic) equation:
\begin{equation}
 \mu\left(\frac{|\vec{a}|}{a_0}\right) \vec{a} = \vec{g_N} 
\end{equation}
where
\begin{equation}
 \mu(x) = \left\{ \begin{array}{l} x \; if \; x << 1\\
                  1 \; if \; x>>1 \end{array} \right.
\end{equation} 
Here $\vec{g_N}$ is the standard Newtonian acceleration and $a_0$ a constant,
observationally determined to be $a_0 = 1.2 \cdot 10^{-8}$ cm/sec$^2$ \citep{b91, sm02}.
The above formulation of MOND conserves angular momentum and energy only for spherical
mass configurations, but a more general formulation exists which obeys 
these conservation laws for all cases \citep{bm84}.

Predictions from MOND have been shown to be in good agreement with
the observed rotation curves of galaxies \citep{b91, s96, sh98} and can also 
explain the velocities of galaxies in groups with reasonable $M/L$ values for
individual galaxies \citep{m02}. In addition, \citet{m95} showed that the velocity
dispersion of 7 dwarf galaxies which was available at the time was compatible 
with the predictions from MOND, a conclusion also found by \citet{l02} with
updated data for the Fornax and Draco dwarf galaxies. These successes
are remarkable given the fact that unlike the dark matter hypothesis, MOND has only
one free parameter ($a_0$) that can be adjusted to explain observations. In addition,
\citet{b04} has recently developed a relativistic formulation of MOND, putting the
theory on a more solid theoretical basis.

Part of the trouble in deciding whether MOND or dark matter is the better candidate
in explaining the velocities of stars on galactic scales stems from the fact that 
cosmological structure formation is still not sufficiently understood,
so within certain limits, the mass and size distribution of dark matter can be adjusted to
fit the observational data. It is therefore highly desirable to test MOND
for objects in which no dark matter is thought to exist. An ideal candidate for such
objects are globular clusters, which still form today as a result of collisions between gas clouds 
during major mergers of galaxies \citep{ws95}, and which are
believed to have formed in the same way in the early universe \citep{az92}.

Testing gravity for low accelarations by using nearby globular clusters was already 
tried by \citet{sc03}. However, nearby clusters experience an acceleration from the Milky Way that
is larger than the critical MOND constant $a_0$ and should therefore by governed
by Newtownian dynamics if MOND is correct. 

In the present paper we investigate the effects of MOND
for a number of low-mass globular clusters in the outer halo of the Milky Way. It will
be shown that if MOND is true, these clusters would have mass-to-light ratios far
larger than their Newtonian values, allowing an independent test of the predictions of 
MOND. The paper is organised as follows: In section 2 we calculate expected velocity
dispersions and mass-to-light ratios for a number of globular clusters for the Newtonian 
case and in the MOND regime. Section 3 discusses how the validity of MOND could be
constrained if the velocity dispersions of these clusters 
would be known and in section 4 we will draw our conclusions.

\section[]{Evaluating the effect of MOND for globular clusters}

\subsection{LOS velocity dispersions for different cases}

In case of Newtonian gravity, the virial theorem connects the mass $M_C$, radius $r$ and 
average velocity dispersion $\sigma$ of a cluster through the following 
equation \citep{bt86}, eq. 4-80a:
\begin{equation}
 \sigma^2 = \frac{G M_C}{r_v} \;\; ,
\end{equation}
where $r_v$ is the virial radius, which in many stellar systems can be 
approximated by the three-dimensional half-mass radius $r_h$ as $r_v \approx 2.5 r_h$ 
if the clusters are stationary systems and sufficiently unperturbed 
that the assumption of virial equilibrium is valid.
A similar relation exists between the three-dimensional half-mass radius and the easier to observe
two-dimensional, projected half-mass radius $r_{hp}$: $r_{hp} = \gamma r_h$ with $\gamma \approx 0.74$.

If we assume an isotropic velocity distribution, the line-of-sight velocity
dispersion is related to the three-dimensional one through $\sigma_{LOS} = \sigma/\sqrt{3}$,
and we obtain the following equation
for the observed velocity dispersion of a star cluster in case of Newtonian gravity
\begin{equation}
 \sigma_{LOS, N} = 0.335 \sqrt{\frac{G M_C}{r_{hp}}} \;\; .
\label{slosn} 
\end{equation}

In order to test the accuracy of eq.\ \ref{slosn}, we created $N$-body representations
of \citet{king66} models
with dimensionless central concentrations in the range $3 \le W_0 \le 15$ and checked
our theoretical estimate against the $N$-body data. King models with concentrations
in this range are usually used to represent density profiles of globular clusters.
Fig.~1 compares the average velocity dispersion of stars in these models with the
prediction of eq.\ \ref{slosn}. For all models, the agreement is within 5\%, which
is accurate enough for our purpose of predicting velocity dispersions in real clusters.
\begin{figure}
\epsfxsize=8cm
\epsffile{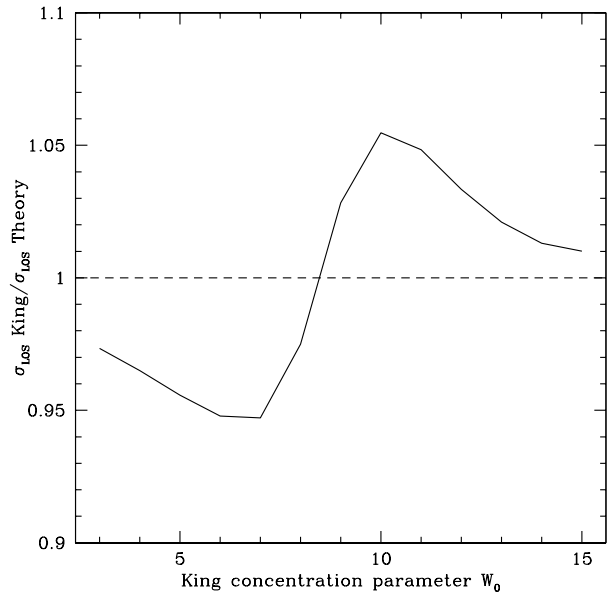}
\caption{Ratio of the line-of-sight velocity dispersion of stars in different
 King models with the prediction from eq.\ \ref{slosn} (solid line). For all models, the 
  velocity dispersion agrees with the prediction to within 5\%.}
\end{figure}

In case of MOND, solutions exist only for several special cases since the function $\mu(x)$ is
defined only for limiting values of $x$. If the acceleration $a$ that stars experience 
is much larger than $a_0$, $\mu \approx 1$ and the MOND solution is the same as in the 
Newtonian case. For this to be true it does not matter if the acceleration $a$ is the internal 
acceleration $a_{int}$ due to stars
in a cluster or if it comes from an external gravitational field $a_{ext}$.
External accelerations are for example
important for all globular clusters which have galactocentric distances $R_{\rm GC} < 
{\rm few} \cdot 10$ kpc
or for the dynamics of the solar system. In both cases the external acceleration due to the Milky 
Way is larger than $a_0$
and the dynamics can be described by Newtonian gravity, no matter how small the internal accelerations
are. Hence, MOND can be important only for
star clusters in the outer halo of the Milky Way.

If both $a_{int}$ and $a_{ext}$ are much smaller than $a_0$ one is in the deep MOND regime with
$\mu(x) \approx x$. If in this case the external acceleration is larger than the internal one
$a_{ext} >> a_{int}$, the system is again nearly Newtonian but with an effective gravitational
constant $G$ that is larger than the standard
Newtonian one by a factor $a_0/a_{ext}$ \citep{m86}. The line-of-sight velocity dispersion is therefore equal to
\begin{equation}
 \sigma_{LOS, M1} = 0.335 \sqrt{\frac{G M_C}{r_{hp}}} \sqrt{\frac{a_0}{a_{ext}}} \;\; .
\end{equation}

\begin{table*}
\caption{Distant globular clusters for which the predictions of MOND and Newtonian dynamics differ. 
 For explanation of symbols see text.}
\begin{tabular}[b]{lcrrrccccrc}
Cluster & lg $M_C$ & $r_{hp}$  & $R_T$  & $R_{\rm GC}$  & $a_{int}$  & $a_{ext}$ & $\sigma_{LOS, N}$ &
$\sigma_{LOS, M}$ &  M/L \\
  Name  & [$M_\odot$] & [pc] & [pc] & [kpc] & [$10^{-9}$ cm/s$^2$] & [$10^{-9}$ cm/s$^2$]  & [km/sec] & [km/sec] &    MOND  \\
\noalign{\smallskip}
\hline
\noalign{\smallskip}
AM 1     & 4.10 & 17.7 & 151.3 &123.2 & $1.84$ & $1.05$ & 0.58 &  1.77 & 18.2 \\
Eridanus & 4.27 & 10.5 & 145.5 & 95.2 & $3.79$ & $1.36$ & 0.93 &  1.96 &  8.8 \\
Pyxis    & 4.52 & 15.6 & 101.1 & 41.7 & $3.38$ & $3.11$ & 1.01 &  2.25 &  9.9 \\
Pal 3    & 4.50 & 17.8 & 173.5 & 95.9 & $2.90$ & $1.35$ & 0.92 &  2.23 & 11.6 \\
Pal 4    & 4.63 & 17.2 & 212.0 &111.8 & $3.48$ & $1.16$ & 1.09 &  2.40 &  9.6 \\
Pal 14   & 4.11 & 24.7 & 103.4 & 69.0 & $1.33$ & $1.88$ & 0.50 &  1.27 & 12.8 \\
Pal 15   & 4.42 & 15.7 &  87.6 & 37.9 & $2.98$ & $3.42$ & 0.90 &  1.68 &  7.0 \\
Arp 2    & 4.34 & 15.9 &  56.3 & 21.4 & $2.68$ & $6.06$ & 0.81 &  1.14 &  4.0 \\
\end{tabular}
\end{table*}

Finally, if $a_{ext} << a_{int}$ and both are smaller than $a_0$, the cluster
is effectively isolated and the acceleration of the cluster stars
is given by $g_M = \sqrt{a_0 g_N}$ where $g_N$ is the acceleration in the Newtonian case.
The line-of-sight velocity dispersion of a star cluster is then given by \citep{m94}: 
\begin{equation}
 \sigma_{LOS, M2} = 0.471 \left(a_0 G M_C \right)^{1/4} \;\; ,
\label{slosm2} 
\end{equation}
i.e. independent of the clusters radius.

\subsection{Application to globular clusters}

Table~1 shows the predicted velocity dispersions of globular clusters for the Newtonian and
the MOND case. We have calculated the dispersions for all galactic globular clusters in the list of
\citet{h96}, but present only those in Table~1 for which there is a noticeable difference between
the two cases. Clusters in Table~1 are generally far away in the galactic halo so that the external acceleration
due to the Milky Way is small, and also have small masses and large half-mass radii so that their internal
accelerations become small. 

Galactocentric distances $R_{\rm GC}$ and projected half-mass radii are taken from \citet{h96}. We have
assumed that for the clusters of Table~1 the half-light radius is equal to the half-mass
radius i.e. mass follows light. This is probably a good assumption since the half-mass relaxation times
of most clusters in Table~1 are of the order of a Hubble time or even larger, so dynamical evolution is not
likely to play an important role for these clusters. Cluster masses were calculated
from the absolute luminosities by assuming a stellar mass-to-light ratio of $M/L=2$. This value is
in agreement with observed mass-to-light ratios of globular clusters \citep{m91,pm93}. \citet{m91} 
for example fitted single-mass
King models to a sample of 32 clusters with reliable central velocity dispersions and
determined an average $M/L = 1.21$ for their sample, while \citet{pm93}
used multi-component King-Michie models for 56 clusters and estimated an average global mass-to-light ratio of 
$M/L=2.3$. Since multi-component models can take the effect of mass segregation into account, the
latter value is probably closer to the truth.

The tidal radii $R_T$ in Table 1 were calculated from the cluster masses by assuming a flat
rotation curve of the Milky Way with $V_G=200$ km/sec for all galactocentric distances. For most clusters 
they are a factor of 5 to 10 larger than the half-mass radii, which means that tidal effects play no significant
role for the internal dynamics of the clusters in our sample. This is a significant improvement compared
to the situation for dwarf galaxies which generally have larger $R_H/R_T$ ratios and for which 
there is an ongoing
discussion to which extent the high-$M/L$ ratios found are caused by tidal effects 
\citep{k97, kk98, o01, fk03, w04}.

Internal accelerations were calculated at the
clusters half-mass radii from $g_M = \sqrt{a_0 g_N}$, assuming that MOND is true, while the external 
accelerations were calculated from
\begin{equation}
a_{ext} = \frac{V_G^2}{R_G}
\end{equation}
and hold in both the Newtonian and the MOND case. The line-of-sight velocity dispersions for the
MOND case were calculated from eqs.\ 5 and 6, depending on whether the internal or external acceleration
is larger. While this procedure is not strictly valid within the framework of MOND, it is accurate enough to give an
estimate of the effect MOND would have. The $M/L$ values in the last column finally are 
calculated from the $\sigma_{LOS, M}$ values assuming that an observer interpretes them for the
Newtonian case. They should be compared with the Newtonian input value of $M/L = 2$ which we assumed. 

\section{Discussion}

From Table~1, it is evident that if MOND is true, $\sigma_{LOS}$ and the deduced $M/L$ are increased by a 
significant 
amount over the Newtonian case. Three clusters would have $M/L$ values in excess of 10, which is
far larger than what is found in any galactic globular cluster. In the analysis of \citet{m91} for example
no cluster had an $M/L$ larger than 3.0 and low-mass clusters with $lg M_C < 4.5$ generally had $M/L < 1.5$.
Similarly, in the analysis of \citet{pm93} low-mass clusters with $lg M_C < 5$ had an average $M/L = 1.8$ and
none had an $M/L$ larger than 3.2.
If a high $M/L$ ratio would be found in any cluster, it could therefore be interpreted 
in only one of three ways:
\begin{enumerate}
\item The initial mass-function of this cluster was radically different from ordinary GCs and heavily weighted 
 either towards high-mass stars which have by now turned into compact remnants,
 or towards very low-mass stars in order to create such a large $M/L$.
\item The cluster contains a significant amount of cold dark matter.
\item Newtonian dynamics has to be modified in the limit of low accelerations.
\end{enumerate}
Possibility (i) can almost certainly be excluded since such high $M/L$ ratios have not been found for
any stellar population to date. In addition, \citet{s99} and \citet{s97} determined ages and metallicities for 
four of the
clusters in our list (Eridanus, Pal 3, 4 and 14) which place them among other galactic globular clusters, 
indicating that there is nothing unusual about these clusters. 

\citet{p84} and others after him have suggested
that globular clusters formed in cold dark matter mini-halos in the early universe, in which case they should 
contain significant amounts of dark matter. So possibility (ii) seems entirely possible and a detection 
of dark matter dominated globular clusters would be a direct confirmation of this idea. In this case
halo globular clusters could help
to bridge the gap between the predicted number of DM sub-clumps in the halos of galaxies and the observed number
of clumps in the form of dwarf galaxies \citep{c02}. Since the formation history and dark matter content would 
basically be the same, halo globular clusters could in such a case be viewed as smaller-sized versions of
dwarf galaxies.

However, observations of interacting and starburst galaxies have 
shown that the formation of globular cluster is still an ongoing process, triggered mainly by major galaxy
mergers \citep{ws95}. Similarly, the relatively high metallicities and younger ages found by \citet{s99} 
and \citet{s97} speak against a primordial formation scenario of these globular clusters and for an accretion
scenario \citep{mg04}. It therefore appears unlikely that the clusters in Table~1 contain a significant amount
of dark matter. 

Another possible interpretation of high velocities would be that Newtons law of gravity has to be changed 
for low accelerations. Observations of additional halo clusters for which MOND does
not predict a velocity enhancement would help to distinguish between the dark matter hypothesis and MOND.
Also, checking the evolutionary state of the clusters which are in the MOND regime could help decide
between MOND and dark matter since the relaxation time in the MOND regime is significantly smaller than 
in the Newtonian case \citep{cb04}, meaning that the clusters would be dynamically
more evolved if MOND is true.

A second possible outcome is that the observed velocity dispersion is in complete agreement with
the Newtonian value and far lower than what MOND predicts. In such a case MOND would probably have to be discarded
in its present form since it would be impossible to reconcile the high rotational velocities and velocity 
dispersions seen in galaxies with the low velocities seen in the clusters of Table~1: Assuming Newtonian
gravity \citet{k96} and \citet{s03} found from an analysis of the orbital motion of galactic satellites
that the total mass of the Milky Way within 50 kpc is respectively $(4.9 \pm 1.1) \cdot 10^{11} M_\odot$ 
and $(5.5 \pm 0.2) \cdot 10^{11} M_\odot$, giving rise to a rotational velocity of $V_G=210$ km/sec.
On the other hand, the mass of the Milky Way in the disc and bulge in the form of visible matter 
is of order $5.5 \cdot 10^{10} M_\odot$ \citep{db98}, giving rise to a rotational velocity of
$V_G = 164 \left( \frac{a_0}{10^{-8} cm/sec^2} \right)^{1/4}$ km/sec within the framework of MOND (eq.\ \ref{slosm2}).
Values of $a_0$ below $5 \cdot 10^{-9}$ cm/sec$^2$ seem therefore to be very hard to reconcile
with the orbital velocities of galactic satellites, even given the uncertainties in the above
numbers. Similarly the galaxies analysed by \citet{b91}
give a mean $a_0 = (1.21 \pm 0.27) \cdot 10^{-8}$ cm/sec$^2$, excluding values of $a_0$ below $6.7 \cdot 10^{-9}$
cm/sec$^2$ at the $2\sigma$ level. On the other hand, even an $a_0$ as low as $5 \cdot 10^{-9}$ cm/sec$^2$ would
still give a velocity dispersion of 1.42 km/sec for AM 1 and 1.80 km/sec for Pal 3, which is twice the Newtonian value.
Hence, velocity dispersion measurements will have the power to falsify or support MOND. 

\section{Conclusions}

We have calculated the velocity dispersions for a number of globular clusters in the halo
of the Milky Way and shown that in case of MOND they would significantly exceed the corresponding
Newtonian values, allowing a test of MOND and dark matter theories for a new class of objects and for
length scales one order of magnitude smaller than where they could be tested before. We showed that 
interesting insights can be obtained
about the formation of globular clusters and the role of dark matter almost independently of the 
actual results, so an observational effort to determine the velocity dispersions for clusters
in our sample would be highly rewarding. Measurements of this kind are feasible with the current 
generation of 8 to 10-m class telescopes.

Compared to dwarf galaxies, the studied globular
clusters also have half-mass radii one order of magnitude smaller than their tidal radii, which
means that they are effectively isolated. Their velocity dispersions are therefore much
more straightforward to interpret since tidal effects are not likely to play an important role.
Also, unlike tests for MOND based on the velocity dispersion of stars in the halos of 
nearby globular clusters, the interaction of the
cluster stars with the galactic tidal field and the proper separation of members and non-members
is not likely to be a problem for our clusters.

\section*{Acknowledgements}

We thank Professor M. Milgrom for many helpful discussions in understanding MOND and the referee,
Luca Ciotti,for comments which improved the presentation of the paper.

\bsp

\label{lastpage}

\end{document}